\documentclass[11pt]{article}

\usepackage{hyperref,algorithmic,algorithm}
\usepackage{amsfonts}
\usepackage{graphicx,subfig}
\usepackage{amsmath,amsthm}
\usepackage{wrapfig}
\usepackage{amssymb}

\usepackage{epsfig}
\usepackage{amsmath}
\usepackage{amssymb}
\usepackage{psfrag}
\usepackage[absolute]{textpos}
\usepackage{setspace}

\newtheorem{thm}{Theorem}[section]
\newtheorem{lem}{Lemma} [section]
\newtheorem{definition}{Definition}

\newcommand{\C}{\mathcal{C}}
\newcommand{\A}{{\bf A}}

\def\x{{\bf x}}
\def\y{{\bf y}}

\def\cc{{\bf c}}
\def\bb{{\bf b}}
\def\aa{{\bf a}}

\newcommand{\beq}{\begin{equation}}
\newcommand{\eeq}{\end{equation}}
\newcommand{\bea}{\begin{eqnarray}}
\newcommand{\eea}{\end{eqnarray}}

\newcommand{\Prob}{\ensuremath{\mathbb{P}}}
\long\def\symbolfootnote[#1]#2{\begingroup%
\def\thefootnote{\fnsymbol{footnote}}\footnote[#1]{#2}\endgroup}

\begin{document}
\title{Summary Based Structures with Improved Sublinear Recovery for Compressed Sensing}

\author{ M. Amin Khajehnejad$^{\dag}$, Juhwan Yoo$^{\dag}$, Animashree Anandkumar$^{*}$  and Babak Hassibi$^{\dag}$ \\ $\dag$California Institute of Technology, Pasadena CA 91125\\ $*$University of California Irvine CA 92697
\thanks{This work was supported in part by the National Science Foundation under grants CCF-0729203, CNS-0932428 and CCF-1018927, by the Office of Naval Research under the MURI grant N00014-08-1-0747, and by Caltech's Lee Center for Advanced Networking.}
}


\maketitle

\begin{abstract}
\boldmath
We introduce a new class of measurement matrices for compressed sensing, using low order summaries over binary sequences of a given length. We prove recovery guarantees for three reconstruction algorithms using the proposed measurements, including $\ell_1$ minimization and two combinatorial methods. In particular, one of the algorithms recovers $k$-sparse vectors of length $N$ in sublinear time $\text{poly}(k\log{N})$,  and requires at most $\Omega(k\log{N}\log\log{N})$ measurements. The empirical oversampling constant of the algorithm is significantly better than existing sublinear recovery algorithms such as Chaining Pursuit and Sudocodes. In particular, for $10^3\leq N\leq 10^8$ and $k=100$, the oversampling factor is between 3 to 8. We provide preliminary insight into how the proposed constructions, and the fast recovery scheme can be used in a number of practical applications such as market basket analysis, and real time compressed sensing implementation.

\end{abstract}


% \fi
\maketitle
%\IEEEpeerreviewmaketitle

\section{Introduction}
\label{sec:Intro}
Despite significant advances in the field of Compressed Sensing (CS), certain aspects of CS remain relatively immature. Thus
far, CS has been viewed primarily as a data acquisition technique~\cite{rice}. As a result, the applicability of CS to other computational applications has not enjoyed commensurate investigation. In addition, to the best of the authors' knowledge, there is no unified CS system that has been implemented for practical real-time applications. A few recent works have addressed the former by applying sparse reconstruction ideas
to certain inference problems including learning and adaptive computational schemes ( e.g. \cite{Devavrat,Cliques,Cevher}). Several other works have addressed the latter by designing hardware, which exploits the fact that CS enables the monitoring of a given bandwidth at a much lower sampling rate than traditional Nyquist-based methods (see e.g., \cite{Eldar2}). The motivating factor behind these works is that for a
given maximum sampling rate (limited by the poor power consumption scaling with sampling rate) achievable by digitizing hardware, it is
possible to either acquire signals over a much greater bandwidth, or with much less power for a given bandwidth. Recent work, inspired by this line of thought, has led to the development of hardware CS encoders (see e.g. \cite{Baraniuk_Hardware,Eldar_Hardware,Lu_Hardware,Juhwan_Hardware}). However, none of the previous works address the problem of real-time signal decoding, which is a critical requirement in many applications.

Although variant by the nature of the problem and physical constraints, perhaps two fundamental issues in the practical implementations
of CS are the following: 1) construction of  measurement matrices that are provably good, certifiable and inexpensive to implement (either
as real time sketches or as pre-built constructions), 2) Time efficient and robust recovery algorithms. Our aim is to introduce and provide an analysis of a sparse reconstruction system that addresses the aforementioned problems and allude to the extensions of CS in the less explored directions.\\
\indent We introduce a new class of measurement matrices for sparse recovery that are deterministic, structured and highly scalable. The constructions are based on labeling the ambient state space with binary sequences of length $n=\log_2{N}$, and summing up entries of $\x$ that share the same pattern (up to a fixed length) at various locations in their labeling sequences. The class of corresponding matrices are RIP-less matrices that are congruent with the Basis Pursuit algorithms, which are standard techniques for sparse reconstruction~\cite{Basis_Pursuit}. In addition, we provide two efficient combinatorial algorithms along with theoretical guarantees for the proposed measurement structures. The proposed algorithms are sub-linear in the ambient dimension of the signal. In particular, we propose a summarized support index inference (SSII) algorithm with a running time of $\mathcal{O}(poly(k\log{N}))$ that requires $\mathcal{O}(k\log{N}\log\log{N})$ measurements to recover $k$-sparse vectors, and has a  empirical required over-sampling factor significantly better than existing sublinear methods. Due to the particular structure of the measurements and decoding algorithms, we believe that the proposed compression/decompression framework is amenable to real time CS implementation, and offers significant simplification in the design of an existing CS encoder/decoder. Furthermore, observations collected based on the proposed constructions appear as low order statistics or ``summaries'' in a number of practical situations in which a similar intrinsic labeling of
the state space exists. This includes certain inference and discrete optimization problems such as market basket (commodity bundle) analysis, advertising, online recommendation systems, genomic feature selection, social networks, etc.\\
\indent It should be acknowledged that there are various results on sublinear sparse recovery in the literature, including \cite{Gilbert,sudocode,Milenkovich,Cormode,Indyk}. Unlike most previous works, the constructions of this paper offer sublinear storage requirement and are compatible with the practical scenarios that we consider. The recovery time of the algorithm is sublinear in the signal dimension, and the empirical recovery bounds are significantly better than the existing sublinear algorithms, such as Chaining Pursuit and Sudocodes, especially for small and moderate sparsity levels and very large signal dimensions.

\section{Proposed Measurement Structures} \label{sec:measurements}

\noindent We define a class of structured binary measurement matrices, based on the following definition
\begin{definition}\label{def:codebook}
Let $m$,$n$ and $d$ be integers. A $(n,d)$ summary is a pair $X=(S,\cc)$, where $S$ is a subset of $\{1,2,\cdots,n\}$ of size $d$, and $\cc$ is a binary sequence of length $d$. A $(m,n,d)$ summary codebook is a collection $\mathcal{C}=\{(S_i,\cc_j)~|~1\leq i\leq m,~0\leq j\leq 2^d-1\}$ of $(n,d)$ summaries, where $S_i$'s are distinct subsets, and $\cc_j$ is the length $d$ binary representation of the integer $j$. If $m={n \choose d}$, $\mathcal{C}$ is called the complete $(n,d)$ summary codebook.
\end{definition}

\begin{figure}[h]
\centering
\includegraphics[width=0.4\textwidth]{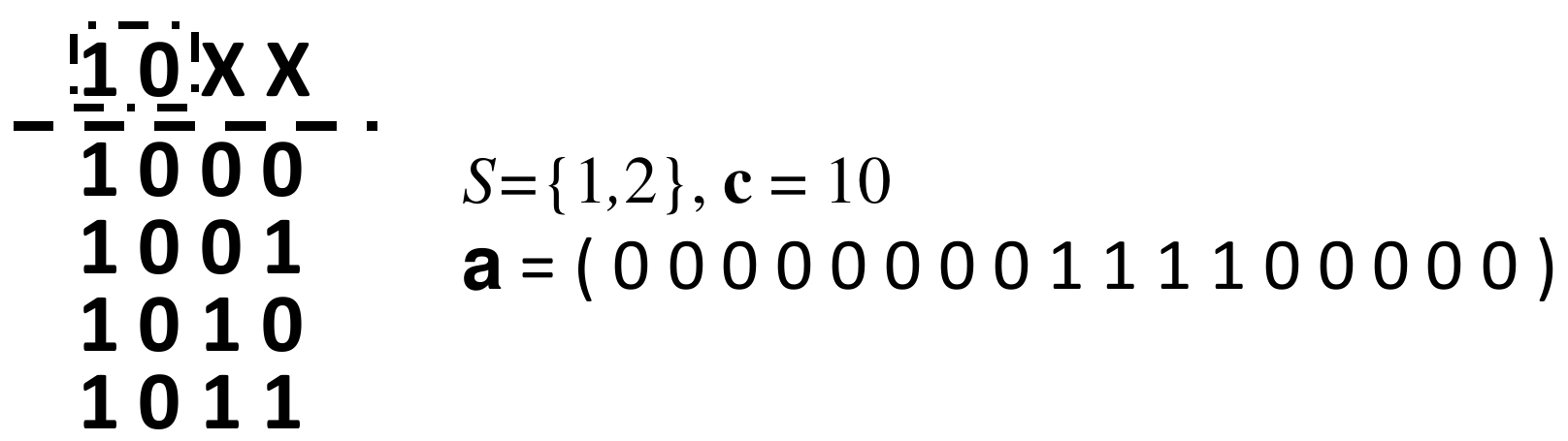}
\caption{\small An example $(4,2)$ summary and the corresponding row of the structured measurement matrix.}
\label{fig:checks}
\end{figure}

To a given $(m,n,d)$ summary codebook $\mathcal{C}$, we associate a binary matrix $A$ of size $M\times N$ where $M = 2^d\times m$, and $N=2^n$, in the following way. For every $(S,\cc)\in \mathcal{C}$, there is a row $\aa=(a_1,\dots,a_N)$ in $A$ that satisfies:
\begin{eqnarray}
a_j = \mathbf{1}\left\{\bb_{j}(S) = \cc\right\}~1\leq j \leq N \label{eq:concide_measurements}
\end{eqnarray}
\noindent where $\bb_{j}$ is the $n$-bit binary representation of $j$, and $\bb_{j}(S)$ is the subsequence of the binary sequence $\bb_{j}$, indexed by the entries of the set $S$\footnote{Note that these structured matrices can be defined for any finite alphabets other than the binary field.}. In other words, $\aa$  has a 1 in those columns $\ell$ whose binary labeling conform to $(S,\cc)$. Every column of $A$ has exactly $m$ ones, and each row has exactly $2^{n-d}$ ones. To clarify this definition, we consider the following example illustrated in Figure \ref{fig:checks}, in which $n=4$ and $d=2$. Suppose that a summary $(S,\cc)$ is given with $S = \{1,2\}$ and $\cc = 10$. All possible binary sequences of length 4 that match $(S,\cc)$ are listed in Figure \ref{fig:checks}. To find the corresponding indices of the listed labels, we should convert them to decimal values and increase by 1, which gives $9,10,11$ and $12$. The row $\aa$ of a measurement matrix that includes this summary is a vector of length $2^4$ that has a $1$ in those indices, as displayed.

%
%\input{CCS_algorithm}
%
%\input{recovery_guarantees}
%
%\input{Simulations}
% Section 0 Background on Bit Flipping and Factor Graphs

% Section 1 Description of the Measurement Matrices
%\input{measurement_structures}

The defined matrices are very well motivated by some practical problems. In general, in a situation where the given signal space retains an intrinsic structured labeling similar to the one described, such constructions prove very useful. In particular, we consider the following two motivational examples.\\
\noindent \textbf{\emph{Resource Optimization.}} Assume that a set $\mathcal{F} = \{F_1,F_2,\cdots,F_n\}$ of features (or parameters) is available, and assume that certain accumulations or collections of features form "lucrative" profiles (structures). In particular, a lucrative profile can be a subset of features which is representable by a binary sequence $\mathbb{\bb}=b_1b_2\dots b_n$, where $b_i$ determines the presence of the $i$'th feature. A practical assumption is that lucrative profiles are limited and weighted, meaning that their profitabilities are variable. The vector $\x = (p_1,p_2,\dots,p_{2^n})^T$ formed by the respective profits of all feature collections is thus an approximately sparse vector. Furthermore, the available information about the profitability of profiles is often  derived from a pool of observations or real world implementations, and are mostly given in the form of summaries. More formally, what can be learned is the average profitability of a certain configuration of only $d$ features. For example, it can be assessed that when $F_1$ and $F_2$ are present and $F_3$ is absent, regardless of all other features, the average profit is some $\overline{p}$.  The collection of summaries form an observation vector $\y$, that is related to $\x$ through a set of linear equations $\y = A\x$, where $A$ has a form similar to those obtained by summary codebooks. This setting arises in many practical applications such as market basket (commodity bundle) analysis, where the objective is to configure the structure of a market that complies the best with the needs and the behaviors of the customers. To that end, it is essential to understand which market configurations are winning and what packages of features (e.g. commodities, pricing options, interest rates, etc.) should be offered to customers, and with what percentages . Furthermore, the customers' behavioral information is often given in terms of high level summaries, e.g. in the lines of the statement ``people who buy A and B, are likely to buy C''.\\
\noindent \textbf{\emph{Compressed Sensing Hardware.}} There are a few factors that severely limit the scalability of the existing CS hardware designs to larger problem dimensions. One of these factors is the generation of the measurement matrix $A$. In the simplest existing design, $A$ is typically a pseudo-random matrix generated with a linear feedback shift register (LFSR)~\cite{Eldar_Hardware,Juhwan_Hardware}. The timing synchronization of a large number of measurements as well as the planar nature of physical implementations is very limiting. Using a more structured matrix may allow considerable simplification and reduction of the required hardware easing some of the previously mentioned limitations. The measurement structure defined in this work is potentially highly amenable to the implementation of practical CS hardware, due to the following two reasons. 1) There exist simple sublinear recovery algorithms for the proposed matrices, other than the linear programming method. This will be elaborated in the proceeding sections.  2) Due to the highly structured design, the integration matrix $A$ can be implemented using one single LFSR seed, and a number of asynchronous digital circuits. Due to the lack of space and the irrelevance of the context, we avoid a detailed description of the latter, and postpone this to a future work.

\section{Proposed Recovery Algorithms}
\label{sec:Methods}
For the measurement matrices described in the previous section we propose three reconstruction algorithms and provide  success guarantees. These algorithms include the Basis Pursuit algorithm (a.k.a $\ell_1$ minimization), as well as two fast algorithms that can recover sparse vectors from a sublinear number of measurements and in a sublinear amount of time. The detailed specifications will be given in the sequel. For the sake of the theoretical arguments that appear in the remainder of this section, we need to define the following notions:

\begin{definition}
\label{def:f_}
Let $n$ and $l$ be integers with $l<n$. We define $f_S(n,l)$, $f_W(n,l,p,\epsilon)$ and $f'_W(n,l,p,\epsilon)$ to be the largest integer $k$ such that when $k$ binary sequences of length $n$ are selected at random, the following happens respectively:
\begin{enumerate}
\item With probability 1, there exists a  $(n,d)$ summary that appears in exactly one of the sequences.
\item With probability at least $p$, for each of the binary sequences, at least a fraction $\epsilon$ of its $(n,d)$ summaries are unique.
\item With probability at least $p$, for each of the binary sequences, at least a fraction $\epsilon$ of its $(n,d)$ summaries that include the first bit are unique.
\end{enumerate}
\end{definition}
\vspace*{-3pt}
It is important to note that the recovery guarantees of the presented combinatorial algorithms are only valid for a class of vectors in which no two disjoint subsets of nonzero coefficients have the exact same sum. For simplicity, we refer to these vectors as ``distinguishable'' signals. This is not the case for  Basis Pursuit.
\vspace*{-4pt}
\subsection{Basis Pursuit}
The success of the basis pursuit algorithm for recovering sparse signals is certified by several conditions. Two major classes of conditions are the Restricted Isometry Property (RIP) and the null space property~\cite{Basis_Pursuit,Null_Space}. It is provable that the measurement structures defined in this paper do not maintain the RIP properties, due to the  existence of columns with fairly large coherence. This however does not discard the suitability of these constructions for $\ell_1$ minimization, since RIP is known to provide a
sufficient condition (see e.g., \cite{BCT10}). Instead, we prove that certain null space conditions hold for the considered class of matrices, and therefore provide a sparse signal recovery bound for $\ell_1$ minimization. We restrict our attention to nonnegative vectors in this case. The reconstruction method is the following program with the additional nonnegativity constraint.
\begin{eqnarray}
&&\text{minimize} \quad {\| \x \|_{1}} \label{eq:l1 min}\\
\nonumber \quad &&\text{subject to}~A\x = \y,~\x\geq 0
\end{eqnarray}

\noindent The performance of the above program was studied for 0-1 matrices in \cite{Amin_Expanders}. In particular, it was shown that a nonnegative vector $\x$ can be recovered from (\ref{eq:l1 min}), if and only if it is the unique nonnegative solution of the linear system of equations, which is stated formally in the following lemma.
\begin{lem}[from \cite{Amin_Expanders}]
Suppose $A\in \mathbb{R}^{m\times n}$ is a matrix with constant column sum, and $\x_0\in\mathbb{R}^{n\times 1}$ is a nonnegative vector. $\x_0$ is the unique solution to (\ref{eq:l1 min}), if and only if $\x_0$ is the unique nonnegative solution to $A\x = A\x_0$.
\label{lem:unique}
\end{lem}

Using the above lemma, we can evaluate the performance of the Basis Pursuit algorithm when used with the presented measurement matrices. The following theorem is fundamental to this analysis.

\begin{thm}[Strong Recovery for Basis Pursuit]
\label{thm:l1 guarantee}
Let $k\leq f_S(n,d-1)$ be an integer, and let $A$ correspond to a complete $(n,d)$ summary codebook. Then every $k$-sparse nonnegative vector $\x$ is perfectly recovered by (\ref{eq:l1 min}).
\begin{proof}
Let $k\leq f_S(n,d-1)$ and let $\x_0$ be a nonnegative $k$-sparse vector. Also, let the $n$-bit binary labels associated to the support set of $\x_0$ be $\bb_1,\bb_2,\dots,\bb_k$. We show that if $A$ corresponds to a complete $(n,d)$ summary codebook, then $\x_0$ is the unique nonnegative solution to $A\x=\A\x_0$. Therefore, by Lemma \ref{lem:unique} it follows that $\x_0$ can be recovered via (\ref{eq:l1 min}). We prove this by contradiction. Suppose that there is another nonnegative vector $\x\neq \x_0$ with $A\x = A\x_0$. Due to the nonnegativity assumption, we may assume that the support sets of $\x$ and $\x_0$ do not overlap. Let the $n$-bit labels of the support set of $\x$ be the binary sequences $\bb'_1,\bb'_2,\dots,\bb'_{\ell}$. From the definition of $f_S(\cdot)$, we can assert that there is a $(n,d-1)$ summary that appears in exactly one of the sequences $\bb_1,\dots,\bb_k$. Let us assume without the loss of generality that the first $d-1$ bits of $\bb_1$ are unique, and that $\bb_1$ is the all zero binary sequence. Therefore, there are at least $n-d+1$ measurements in $\y=A\x_0$ that are equal to the entry of $\x_0$ that corresponds to the label $\bb_1$. These measurements are those that correspond to the summaries
\beq
(\{1,2,\dots,d-1,i\},{\bf{0}}),~ d\leq i\leq n
\eeq
\noindent Since, $A\x = A\x_0$, there must be a nonzero entries in $\x$ with labeling indices that satisfy the above summaries. In particular, without loss of generality assume that the first $d$ bits of $\bb'_1$ are all zero. However, since the support sets of $\x$ and $\x_0$ do not overlap, $\bb'_1$ is different from $\bb_1$ in at least one bit, say $\bb_1(j) \neq \bb'_1(j)$ for some $j>d$. Now consider the summary $(S,\cc) = (\{1,2,\dots,d-1,j\},00\dots01)$, which represent the set of all binary sequences that are zero on the first $d-1$ bits and one on the $j$th bit. Because $A$ corresponds a complete $(n,d)$ codebook, there is a row of $A$ that is based on $(S,\cc)$, and moreover the corresponding value of $\y$ is nonzero, because $\bb'_1$ conforms to $(S,\cc)$. On the other hand this cannot be true when considering the equations $\y=A\x$, because it requires that one of the labels $\bb_1,\dots,\bb_k$ conform to $(S,\cc)$, which cannot be $\bb_1$ (recall that $\bb_1$ is the all zero codeword, whereas $\cc$ includes a 1). The existence of such a label contradicts the assumption that $\bb_1$ is the only label whose $d-1$ first bits are all zero.

\end{proof}
\end{thm}
The complexity of Basis Pursuit is generally polynomial in the ambient dimension of the signal. Specifically, one can implement (\ref{eq:l1 min}) in $\mathcal{O}(N^3)$ operations, without exploiting any of the available structural information of the measurement matrix. Although there are some advantages to Basis Pursuit, such as robustness to noise, its complexity is impractical for problems where $N$ scales exponentially. In these situations, sublinear time algorithms are preferred.
\vspace*{-4pt}
\subsection{Summarized Support Index Inference}
\label{sec:DCSdescription}
The first sublinear algorithm discussed in this subsection is called the summarized support index inference (SSII). The algorithm is based on iteratively inferring the nonzero entries of the signal based on one of the distinct values of $\y$ and its various occurrences. The method is described below.

At the beginning of the algorithm, distinct nonzero values of the observations $\y$ are identified, and are separated from the zero values. Due to the distinguishability  assumption on $\x$, each distinct nonzero value of $\y$ is a sum of a unique subset of nonzeros of $\x$, and can thus be used to infer the position of at least one nonzero entry. The index of a nonzero entry of $\x$ is determined by its unique labeling, which is a binary sequence of length $n$. Therefore, the algorithm attempts to infer all relevant binary sequences. Suppose that a nonzero value of $\y$ is chosen that has $t$ occurrences, say without loss of generality, $y_{1}=y_2=\dots=y_{t}$. Also, let the $(n,d)$ summary which corresponds to the $i$th row of $A$ be denoted by $(S_i,\cc_i)$ (see equation (\ref{eq:concide_measurements})). The algorithm explores the possibility  that  $y_{1},y_{2},\dots,y_{t}$ are all equal to a single nonzero entry of $\x$, by trying to build a binary sequence $\bb$ that conforms to the summaries $\{(S_{i},\cc_{i})\}_{i=1}^t$, i.e., by setting:
\beq
\label{eq:b(S_i)}
\bb(S_i) := \cc_i,~\forall 1\leq i \leq t
\eeq
\noindent If there is a conflict in the set of equations in (\ref{eq:b(S_i)}), then that value of $\y$ is discarded in the current iteration, and the search is continued for other values. Otherwise, two events may occur. If (\ref{eq:b(S_i)}) uniquely identifies $\bb$, then one nonzero position and value of $\x$ has been determined. It is subtracted, measurements are updated and the algorithm is continued. However, there might be a case where only $n_1 < n$ bits of $\bb$ are determined by (\ref{eq:b(S_i)}). In this case, we use the zero values of $\y$ to infer the remaining $n-n_1$ bits in the following way. Let the set of known and unknown bits of $\bb$ be denoted by $S_1$ and $S_2$, respectively. We consider the summaries $(S,\cc)$ which contribute to $A$, and among all, consider all distinct subsets $S$. If there is a subset $S'$ such that among all the measurements corresponding to $(S',\cc)$ where $\cc$ does not conflict with $\bb(S')$, exactly one of them are nonzero, say $(S',\cc')$, then the bits of $\bb$ over $S'\cap S_2$ can be uniquely determined by setting $\bb(S') = \cc'$. This procedure is repeated until either $\bb$ is completely identified, or all possibilities are exhausted. A high level description of the presented  method is given in Alg. \ref{alg:1}, for which  we can assert the following weak and strong recovery guarantees.
 %Suboptimal greedy algorithm for short term multicast precoder design
%~\\

\begin{algorithm}[t]\caption{\small{SSII}}
\begin{algorithmic}[1]
\STATE Repeat until all nonzeros of $\x$ are identified.
\STATE Identify distinct nonzeros of $\y$, exhaust the following:
\vspace*{-0pt}
\STATE Consider all occurrences of a value $y_{\pi(1)}=\dots=y_{\pi(t)}$.
\STATE  Construct a binary sequence $\bb$ by setting $\bb(S_{\pi(i)}):=\cc_{\pi(i)},~\forall 1\leq i\leq t$, where $(S_j,\cc_j)$ is the summary corresponding to measurement $\y_j$.
\STATE If $\bb$ is fully characterized without confliction from previous step, then a nonzero entry of $\x$ has been determined. subtract it, update $\y$ and go to step 2.  Otherwise, exhaust the following step.
\STATE Find a  subset $S'$, such that among summaries $(S',\cc)$ that do not contradict with $\bb$, exactly one corresponds to a nonzero of $\y$, say $(S',\cc')$, and set $\bb(\cc') := S'$.
\end{algorithmic}
\label{alg:1}
\end{algorithm}
\vspace*{-3pt}
\begin{thm}[Strong Recovery for SSII]
Let $k\leq f_S(n,d-1)$ be an integer, and let $A$ correspond to a complete $(n,d)$ summary codebook. Then every $k$-sparse distinguishable vector $\x$ is perfectly recovered by Alg. \ref{alg:1}.
\begin{proof}
Let $k\leq f_S(n,d-1)$ and let $\x$ be a $k$-sparse vector. Also, let the $n$-bit binary labels associated to the support set of $\x$ be $\bb_1,\bb_2,\dots,\bb_k$. We show that at least one of these labels can be inferred from  one of the nonzero values of the vector $\y=A\x$, by solving (\ref{eq:b(S_i)}). From the definition, there is a $(n,d-1)$ summary that appears in exactly one of the labels $\bb_1,\bb_2,\dots,\bb_k$. Without loss of generality, let's assume that the first $d-1$ bits of $\bb_1$ are unique, and that $\bb_1$ is the all zero binary sequence. Also, let the nonzero value of $\x$ in the position given by $\bb_1$ be $\gamma$.  Now consider all summaries $(S,\cc)$ for which the value of the corresponding entry in $\y$ is equal to $\gamma$. Let these summaries be denoted by $\{(S_i,\cc_i)\}_{i=1}^t$, where $t$ is the number of occurrences of $\gamma$ in $\y$. We show that there is a unique binary sequence $\bb'$ that conforms to all of these summaries. In other words, we prove that equation (\ref{eq:b(S_i)}) has a unique solution which is equal to $\bb' = \bb_1$.

Due to the distinguishability assumption on the nonzero values of $\x$, The set $\{(S_i,\cc_i)\}_{i=1}^t$ should include the following summaries:

\beq
(\{1,2,\dots,d-1,i\},{\bf{0}}),~ d\leq i\leq n
\eeq
\noindent Where ${\bf{0}}$ indicates the all zero bit sequence of length $d$.  Clearly the only length $n$ binary sequence that conforms to all of the above summaries is the all zero binary sequence, namely $\bb_1$. Thus, we only need to show that $\bb_1(S_i)=\cc_i$ for all other summaries $(S_i,\cc_i),~1\leq i\leq t$. This also follows immediately from the distinguishability assumption on $\x$, and the fact that  every instance of $\gamma$ in the vector $\y$ is only the result of the nonzero value in $\x$ labeled by $\bb_1$ (i.e. it is not the direct sum of another subset of the entries of $\x$).
\end{proof}
\end{thm}

\vspace*{-8pt}
\begin{thm}[Weak Recovery for SSII]
\label{thm:DCS guarantee}
Let $k\leq f'_W(n,d,p,\epsilon)$ be an integer, and let $A$ correspond to a random $(n,m,d)$ summary codebook. Then, a random $k$-sparse distinguishable vector $\x$ is recovered by Alg. \ref{alg:1} with probability at least $1-kn\left(1-p+p(1-\frac{\epsilon d}{n})^m\right)$.
\begin{proof}
We define an event $\mathcal{E}$ which is stronger that the success event of Algorithm \ref{alg:1}, namely a sufficient condition for the success of SSII. Let the $n$-bit binary labels associated to the support set of $\x$ be $\bb_1,\bb_2,\dots,\bb_k$, and let $\C$ be the $(n,m,d)$ summary codebook based on which $A$ is constructed. The sufficient condition for success of SSII is that for every $1\leq i\leq k$, and every bit $1\leq j \leq n$, there exists a summary $(S,\cc)$ in $\C$ such that $j\in S$ and in addition $\bb_i(S) = \cc$ and $\bb_{\ell} \neq \cc~\forall \ell\neq i$. In other words, for each of the $k$ labels corresponding to the support of $\x$ and each of the $n$ bits, there is a summary in the codebook that includes the considered bit and only conforms to that particular label.

We find a lower bound on the probability of the complementary event $\mathcal{E}^c$ by using union bounds. Note that there are $m$ distinct subsets in the summaries of the codebook $\C$, which are chosen randomly. We assume that the subsets are chosen independently at random, and allow repetition. In case of repetition, the repeated subset is excluded, which only makes things worst.  Consider a label $\bb_1$ and the first bit. The probability that a randomly chosen subset of bits of length $d$ includes the first bit is $\frac{d}{n}$. Furthermore, let us say that at least a fraction $\epsilon'$ of the summaries that conform to $\bb_1$ and include the first bit, does not conform to the remaining $\bb_i$'s (i.e. only appear in $\bb_1$). Then, when a random subset $S$ is chosen, with probability at least $\frac{\epsilon' d}{n}$, the following happens:

\beq \label{eq:f_event} 1\in S ~\text{and}~ \bb_i(S)\neq \bb_1(S)~\forall 1<i\leq k \eeq
 
\noindent Therefore, the probability that the above event does not happen for any of $m$ randomly chosen subsets $S$ is at most $(1-\frac{\epsilon' d}{n})^m$. From the definition of $f'_W(\cdot)$ and the fact that $k\leq f'_W(n,d,p,\epsilon)$, we know that with probability at least $p$, $\epsilon' \geq \epsilon$, and therefore, the probability that (\ref{eq:f_event}) does not happen for any set $S$ in the codebook $\C$ is at most  $1-p + p(1-\frac{\epsilon d}{n})^m$. If we union bound the probability of such event for all possible $k$ labels and all possible $n$ bits, we conclude that the probability of the undesirable event $\mathcal{E}^c$ is bounded by:
\beq \Prob(\mathcal{E}^c) \leq nk(1-p+ p(1-\frac{\epsilon d}{n})^m) \eeq
\noindent Which concludes the proof of the theorem.    
\end{proof}
\end{thm}
\vspace*{-3pt}
The explicit recovery bounds given by above theorem are calculated in Section \ref{sec:bounds}. Alg. \ref{alg:1} can be implemented very efficiently, with $\mathcal{O}(\max(\text{poly}(M),k\log{N}))$ operations, which is sublinear in the dimension of the problem. The computational advantage is owed to the most part to the structural definition of the measurement matrices which facilitates sublinear search over the column space of the matrix. In addition, we do not require an exponential memory for decoding, since the information about $A$ and the current inferred indices of the unknown vector at each stage can be retained by only storing the corresponding binary indices. %This is a %significant advantage over the Sudocode constructions of \cite{sudocode}, which are unstructured, and require frequent %references to a dense matrix with $\mathcal{O}(N\log{N})$ entries.

\subsection{Mix and Match Algorithm}
\label{sec:CCSdescription}
We describe a third recovery method, which is on the lines of the algorithm proposed in \cite{Devavrat} with slight  modifications. The algorithm consists of two subroutines: a value
identification phase in which the nonzero values of the unknown signal is determined, and a second phase for identifying the support set of $\x$. The method is based on measurements given by $\y = (\y^{(1)T},\y^{(2)T})^T = (A_1^T, A_2^T)^T\x$, where only $\y^{(1)}$ is used for the first phase, and $\y^{(2)}$  and $A_2$ are used in the second phase. For details of this method please refer to \cite{Devavrat}. We analyze this algorithm for the proposed measurement structures of this paper, which is different from the analysis of \cite{Devavrat}.

\begin{algorithm}
\caption{\small{M\&M }}
%Suboptimal greedy algorithm for short term multicast precoder design
%~\\
\begin{algorithmic}[t]
\STATE Find the set $Y$ of nonzero entries of $\y_1$ and set $X=\emptyset$. $X$ will determine the set of nonzeros of $\x$. Repeat steps 2,3 until $S(X) = Y$.
\STATE Update  the set $S(X)$ of sums of subsets of $X$.
\STATE Find the smallest entry of $Y$ that is not in $S(X)$, and add it to $X$.
\STATE Initiate zero binary sequences $\{\bb_x|x\in X\}$, which will determine the labeling of the support indices of $\x$.
\STATE For every nonzero entry of $\y_2$, find the corresponding summary $(S,\cc)$, and a subset $X'\subset X $ that sum up to that value of $\y_2$.  Set $\bb_x(S) = \cc, \forall x\in X' $.
 \end{algorithmic}
\label{alg:2}
\end{algorithm}

\vspace*{-8pt}
\begin{thm}[Weak Recovery for M\&M]
\label{thm:m&m_guarantee}
Let $k\leq f_W(n,d,p,\epsilon)$ be an integer, and $A=\left(A_1^T A_2^T\right)^T$ where $A_1$ and $A_2$ correspond to a random $(m,n,d)$ summary codebook, and a complete $(n,1)$ summary codebook, respectively. Then, a random nonnegative $k$-sparse distinguishable vector $\x$ is recovered by Alg. \ref{alg:2} with probability at least $p\left(1-k(1-\epsilon)^m\right)$.
\begin{proof}
Let the $n$-bit binary labels associated to the support set of $\x$ be $\bb_1,\bb_2,\dots,\bb_k$, and let $\C_1$ be the $(n,m,d)$ summary codebook based on which $A_1$ is constructed. It can be shown that the value identification subroutine of Alg. \ref{alg:2} identifies all nonzero values of the nonnegative vector $\x$ correctly, if in the observation vector $\y$, all every nonzero value of $\x$ appear at least once. We find the probability that this condition holds, when the $m$ subsets of the random coodbook $\C_1$ are chosen at random. For every $1\leq i\leq k$, we define the following set of subsets of $\{1,2,\dots,n\}$:
\beq \mathcal{U}_i\{S~|~ |S|=d,~\bb_j(S)\neq \bb_i(S)~\forall j\neq i\}\eeq
 
\noindent If a subset $S$ in the codebook $\C_1$ belongs to $\mathcal{U}_i$, then the nonzero entry $\gamma_i$ that corresponds to the label $\bb_i$ appears in the observation vector $\y$. Therefore, we are interested in finding the probability  that the set of $m$ subsets of $\C_1$  has a nonempty overlap with all $\mathcal{U}_i$'s. Let us assume that for some $\epsilon'>0$, the following holds:

\beq |\mathcal{U}_i| \geq \epsilon'{n\choose d},~\forall 1\leq i\leq k\eeq 

\noindent When a subset $S$ is chosen at random, the probability that it belongs to $\mathcal{U}_i$ is at least $\epsilon'$. Therefore the probability that $\mathcal{U}_i$ does not overlap with the set of all subsets $S$ appearing in $\C_1$ is at most $(1-\epsilon')^m$. Using a union bound over all $1\leq i\leq k$,  we conclude that the probability that this undesirable event happens for at least one of the sets $\mathcal{U}_i$ is at most $k(1-\epsilon')^m$, which means that the probability of success is at least $1-k(1-\epsilon')^m$. However, we know from the definition of $f_W(\cdot)$, and the fact that $k\leq f_W(n,d,p,\epsilon)$, that with probability at least $p$, we have $\epsilon'>\epsilon$. Therefore the overall probability of success  is at least:
\beq 1-p+p(1-k(1-\epsilon)^m) \geq p(1-k(1-\epsilon)^m)\eeq 
\end{proof}
\end{thm}
\vspace*{-4pt}
The complexity of Alg. \ref{alg:2} is $\mathcal{O}(\max(\text{poly}(M),2^k))$, and thus explodes when $k$ grows.
%
%The value identification subroutine is based on a measurement matrix $A_{1}$ generated by a
%random $(m,n,d)$ concise codebook, where $n=\log_2{N}$, and the choice of $d$ will be explained later.
%The number of linear measurements is therefore $M_1=m\cdot 2^d$. The support identification subroutine  uses the identified values, as well as observations made by a measurement matrix $A_2$ that corresponds to a complete $(n,1)$ concise codebook, namely $\mathcal{C}_2 = \{(i,c)~|~1\leq i \leq n,~c\in\{0,1\}\}$. Therefore, rows $2i-1$ and $2i$ of the measurement matrix $A_2$ correspond to the codewords $(i,0)$ and $(i,1)$, respectively. The V.I. subroutine returns the $n$-bit binary indices of the support set of $\x$, by looking at each entry of the observation vector $\y^{(2)}=A_2\x$, and finding a (unique) subset of the elements of $V$, that sum up (or is as close as possible) to the considered entry of $\y^{(2)}$. This identifies one bit value in the binary index of the corresponding subset of the support set.
\vspace*{-3pt}
\section{Recovery Bounds}
\label{sec:bounds}
We derive recovery bounds for (\ref{eq:l1 min}) and Alg.'s \ref{alg:1} and \ref{alg:2} by  obtaining explicit bounds on the terms of definition \ref{def:f_} and replacing them in the recovery guarantees of Section \ref{sec:Methods}, namely Theorems \ref{thm:l1 guarantee}-\ref{thm:m&m_guarantee}. The proof  of the following lemma is based on some combinatorial techniques and Chernoff concentration bounds. 

\begin{lem}
\label{lem:f_}
Let $n,l$ and $k$ be integers and $0<\alpha <1/2$. Also, let $\epsilon =1- k{ \frac{n}{2}(1+\sqrt{2\alpha})\choose l}/{n\choose l}$, and $p=1-k^2e^{-\alpha n}$. Then,
\begin{enumerate}
\item $f_S(n,l) \geq 2^l$.
\item $f_W(n,l,p,\epsilon) \geq k$.
\item $f'_W(n,l,p,\epsilon) \geq f_W(n-1,l-1,p,\epsilon)$.
\end{enumerate}
\end{lem}

By exploiting the expressions of the above lemma in Theorems \ref{thm:l1 guarantee}-\ref{thm:m&m_guarantee}, we obtain the following bounds for different methods:

\noindent \textbf{\emph{Basis Pursuit}}. If a complete $(n,d)$ summary codebook is used to build $A$, then the number of measurements is $M=2^d{n \choose d}$, and every sparsity $k \leq 2^{d-1}$ is guaranteed to be recovered. When put together (recall that $n=\log_2{N}$), an upper bound on the the required number of measurements for recoverable sparsity $k$ is given by:
\vspace*{-5pt}
\beq M = 2k{\log{N} \choose \log{k}} \eeq
\noindent In particular, for small values of $k$, the above bound is comparable with the $M=2k\log{N}$ bound of $\ell_1$ minimization for random Gaussian matrices~\cite{Venkat}.

\noindent \textbf{\emph{SSII Algorithm.}}
We focus on the weak bound, namely the one obtained from Theorem \ref{thm:DCS guarantee}. The general strategy is to take the values of $p$ and $\epsilon$  according to Lemma \ref{lem:f_} with $l=d-1$, and choose $k$ and $m$ in such a way that firstly, $\epsilon$ is bounded away from zero, and secondly, the probability of recovery failure approaches zero as $n\rightarrow \infty$. Taking $k = \lambda 2^{-d\log_2(\sqrt{\alpha/2}+1/2)}$ for some $0<\lambda <1$, a few basic algebraic steps lead to the following:
\vspace*{-2pt}
\beq \Prob\left(\text{\emph{failure}}\right) \leq k^3ne^{-\alpha n}+kn\left(1-(1-\lambda) d/n\right)^m, \eeq
\vspace*{-1pt}
\noindent  It follows that the above expression approaches zero if $m = \Omega(n\log{n})$. Furthermore, $\alpha$ can be chosen arbitrarily close to zero. Therefore, it follows that an upper bound on the required number of measurements for successful recovery with high probability is given by:
\vspace*{-3pt}
\beq M = \Omega(k\log{N}\log\log{N}).\eeq

\noindent \textbf{\emph{M\&M Algorithm.}}
We take $k = \lambda 2^{-d\log_2(\sqrt{\alpha/2}+1/2)}$, and $\epsilon,p$ according to Lemma \ref{lem:f_}, it follows that:
\vspace*{-3pt}
\beq \Prob\left(\text{\emph{failure}}\right) \leq k^2e^{-\alpha n}+k\lambda^m ,\eeq
\vspace*{-2pt}
\noindent Which asymptotically vanishes if $m = \Omega(\log{k})$. Recall that the number of measurements in this case is determined by the matrix $A = [A_1^T,A_2^T]^T$ described in Theorem \ref{thm:m&m_guarantee}, which is equal to $M=2\log{N} + m\times2^d$. Therefore, it follows that an upper bound on the required number of measurements for successful recovery with high probability is given by:
\vspace*{-3pt}
\beq M = 2\log{N}+\Omega(k\log{k}).\eeq
In particular, when $k=o(\log\log{N})$, this means only $\mathcal{O}(\log{N})$ measurements are required, and the running time of the algorithm is $\mathcal{O}(\log{N})$ (see Section \ref{sec:Methods}), both of which are almost optimal.
\vspace*{-10pt}
\section{Simulations}
\label{sec:Simulations}
\begin{figure}[h]
\centering
\includegraphics[width=0.6\textwidth]{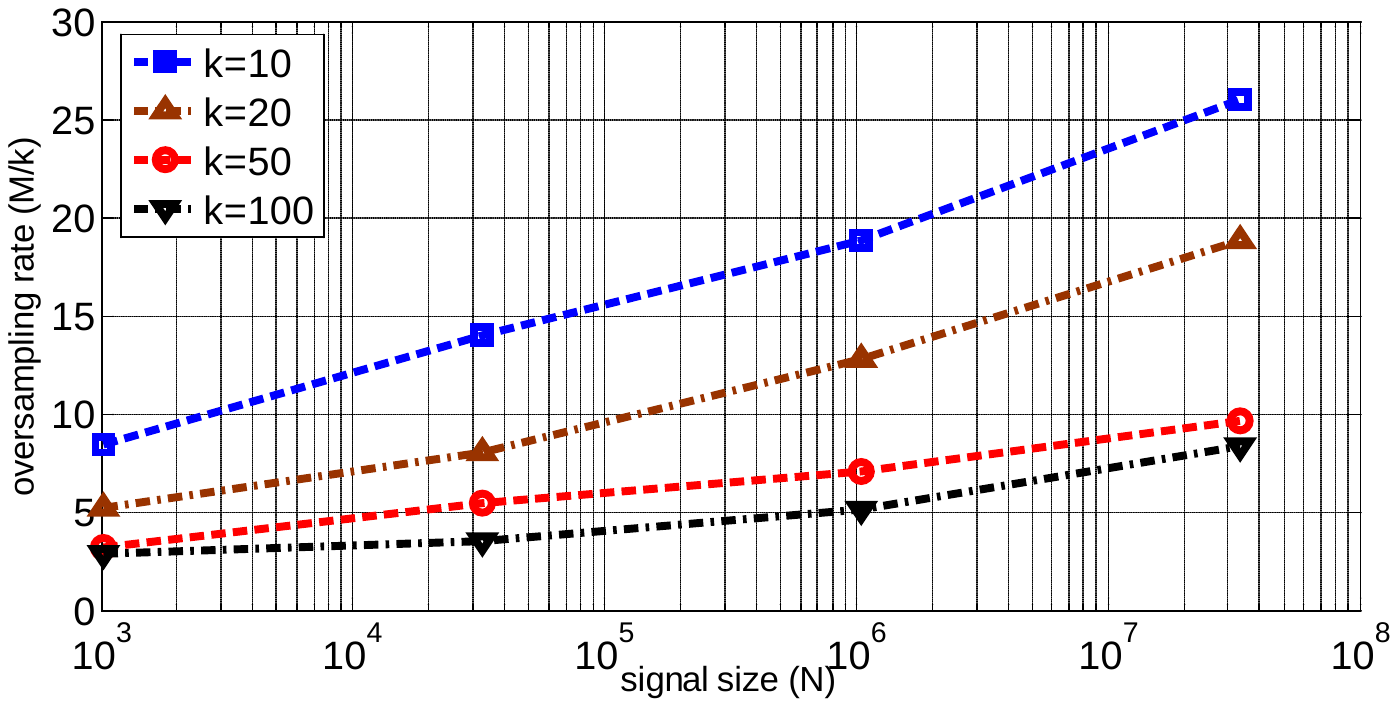}
\caption{\scriptsize Required oversampling rate for successful recovery of Alg. \ref{alg:1} on proposed constructions versus signal dimension for various sparsity levels.}
\label{fig:overS}
\end{figure}

\vspace*{-4pt}
Since Alg. \ref{alg:2} is only efficient for very small values of $k$, we present the empirical performance of Alg. \ref{alg:1}. Due to the efficiency of the method, it is possible to perform simulations for very large values of $N$. In Figure \ref{fig:overS}, the empirical required over-sampling rate for Alg. \ref{alg:2} and the proposed constructions is plotted versus the signal dimension $N$, for various sparsity levels $k$. The required criteria here is that the probability of successful recovery be larger than 90\%. Note that when  $N$ is increased by $3$ orders of magnitude, the required number of measurements is increased by a factor of 3, which is an indication of the logarithmic dependence of $M$ to $N$. Furthermore, as the signal becomes less sparse (i.e. $k$ increases), the required oversampling factor decreases. For $k=100$, this ratio is only about 3 for $N=1024$, and about 8 for $N=3.3\times 10^7$. This is significantly better than existing sublinear recovery algorithms. Note that the optimal value of $d$ for constructing the measurement matrices for every $k,N$ is found empirically.

\begin{figure}[h]
\centering
\includegraphics[width=0.6\textwidth]{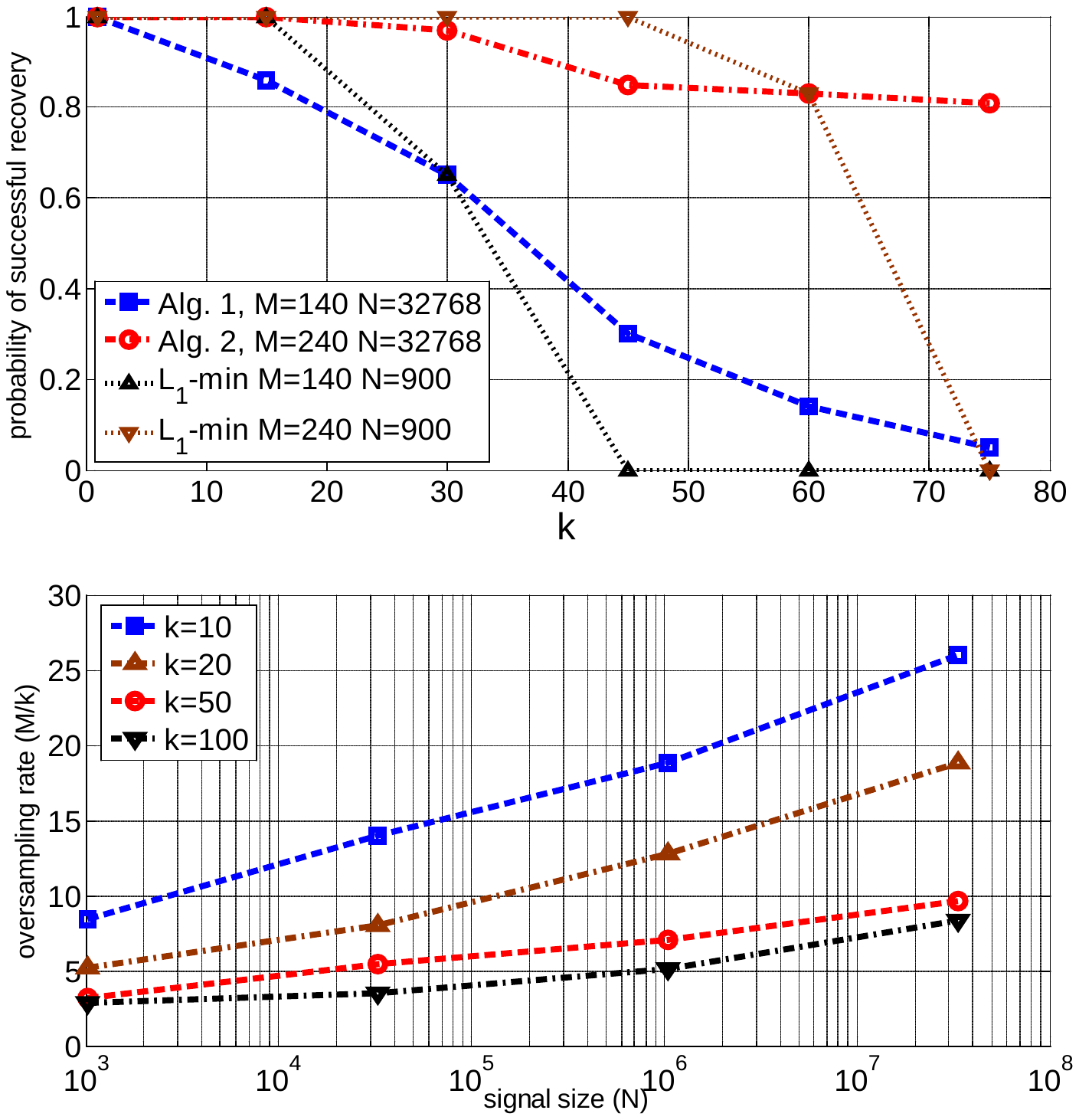}
\caption{\scriptsize Probability of successful recovery of Alg. \ref{alg:1} versus sparsity level $k$, for $N=32768$ and $M=140, 240$, and the same curves for $\ell_1$-minimization over i.i.d Gaussian matrices with $N=900$.}
\label{fig:probR}
\end{figure}

In Figure \ref{fig:probR}, the probability of successful recovery is plotted against the sparsity level $k$ for $N=32768$, and  $M=140$ and $240$. We can see that although the number of measurements has only increased by a factor of $1.7$, the recoverable sparsity (given a fixed probability of success) has improved in some cases by a factor of $5$. These curves are comparable with the performance of $\ell_1$ minimization over dense matrices, with $N=900$, as displayed, which is an indication of the strong performance of the proposed scheme.

%\section{Conclusion}

%\hfill %January 11, 2007

%\appendix{}
%\section{}
%%\label{App:}


\vspace*{-10pt}
\begin{thebibliography}{1}
\vspace*{-5pt}
\bibitem{rice}
Compressed Sensing Online Resources, \url{http://dsp.rice.edu/cs}.
%
%\bibitem{OMP}
%J. Tropp and A. Gilbert.,{\em Signal recovery from random measurements via Orthogonal Matching Pursuit}
%IEEE Trans. Info. Theory, 53(12),4655-4666, 2007.




%\bibitem{Cormode_Algorithmic}
%G. Cormode and S. Muthukrishnan, {\em Towards an Algorithmic Theory of Compressed Sensing}, Tech. Report 2005.
%
%\bibitem{Cormode_Countmin}
%G. Cormode and S. Muthukrishnan, {\em Improved Data Stream Summaries: The count-min sketch and its applications}, FSTTCS 2004.


\bibitem{Devavrat}
S.  Jagabathula and D. Shah, {\em Inferring popular rankings under constrained sensing}, NIPS, 2008.



\bibitem{Cliques}
X. Jiang, Y. Yao and L. Guibas, {\em Stable Identification of Cliques with Restricted Sensing}, NIPS 2009.

\bibitem{Cevher}
V. Cevher, {\em Learning with Compressible Priors}, NIPS 2009.

\bibitem{Null_Space}
Mihailo Stojnic, Weiyu Xu and Babak Hassibi,  {\em Compressed Sensing - Probabilistic Analysis of a Null-space Characterizatio}, ICASSP 2008.

\bibitem{Eldar2}
Y. Eldar, {\em Compressed Sensing of Analog Signals in Shift-Invariant Spaces}, IEEE Tran. on Sig. Proc., 57(8), 2986-2997.
%-----------------These references should all appear as a block:-------------from here

\bibitem{Baraniuk_Hardware}
J. Laska, S. Kirolos, M. Duarte, T. Ragheb and  R. Baraniuk and Y. Massoud, {\em Theory and implementation of an
analog-toinformation coverter using random demodulation}, ISCAS 2007.

\bibitem{Eldar_Hardware}
M. Mishali and Y. C. Eldar, {\em From Theory to Practice: Sub-Nyquist
Sampling of Sparse Wideband Analog Signals}, IEEE J. of Sel.
Top. on Sig. Proc., 4(2):375–391, 2010.

\bibitem{Lu_Hardware}
J. Luo, Yi Lu and B.Prabhaka, {\em Prototyping Counter Braids on NetFPGA}, Tech. Rep. 2008.

\bibitem{Juhwan_Hardware}
J. Yoo, S. Becker, E. Cand\`{e}s and Azita Emami, {\em A Random Modulation Pre-Integration Receiver
for Sub-Nyquist Rate Signal Acquisition}, Preprint 2010.


%----------------------------------------------------------------------------to here
\bibitem{Basis_Pursuit}
E. J. Cand\`{e}s and T. Tao, {\em Decoding by linear programming}, IEEE Trans. Inform. Theory, 51 4203-4215.

\bibitem{Gilbert}
A. Gilbert, M. Strauss, J. Tropp, and R. Vershynin, {\em One Sketch for All: Fast Algorithms for
Compressed Sensing}, STOC 2007.

\bibitem{sudocode}
D. Sarvotham, D. Baron, and R. Baraniuk, {\em Sudocodes - Fast Measurement and Reconstruction of
Sparse Signals}, ISIT 2006.

\bibitem{Milenkovich}
W. Dai, O. Milenkovic and H.Pham, {\em Structured sublinear compressive sensing via dense belief propagation}, Preprint 2011.

\bibitem{Cormode}
G. Cormode and S. Muthukrishnan, {\em Combinatorial algorithms for compressed sensing}, CISS 2006.

\bibitem{Indyk}
R. Berinde, A. Gilbert, P. Indyk, M. Karloff, and M. Strauss, {\em Combining Geometry and Combinatorics: a Unified Approach to Sparse Signal rRecovery},  Allerton 2008.

%-----------------------------------------------------------------------------
\bibitem{BCT10}
J. Blanchard, C. Cartis, and J. Tanner,
{\em Compressed Sensing: How Sharp Is the Restricted Isometry Property?}, SIAM Rev. 53 105-125, 2010.

%
%\bibitem{Sidhant}
%F. Parvaresh, H. Vikalo, S. Misra, and B. Hassibi, {\em Recovering Sparse Signals Using Sparse
%Measurement Matrices in Compressed DNA Microarrays}, IEEE  Jour. of Selected Topics in Sig. Proc., 2(3), 275-285, 2008.

%\bibitem{Weiyu_Expanders}
%W. Xu and B. Hassibi, {\em Efficient Compressive Sensing with Deterministic Guarantees Using Expander Graphs},
%Proc. of IEEE Info. Theory Workshop 2007.


\bibitem{Amin_Expanders}
A. Khajehnejad, W. Xu, A. G. Dimakis, B. Hassibi, {\em Sparse Recovery of Positive Signals with Minimal Expansion}, IEEE Tran. on Signal Proc., 59(1),196-208, 2010.

\bibitem{Venkat}
V. Chandrasekaran, B. Recht, P. Parrilo and A. Willsky, {\em The Convex Geometry of Linear Inverse Problems}, available at arXiv:1012.0621v1.


\end{thebibliography}
\end{document}